\documentclass[conference, 10pt]{IEEEtran}

\usepackage{csquotes}
\usepackage{textgreek}
\usepackage{amsmath}
\usepackage{multirow}
\usepackage{multicol}
\usepackage{soul}
\usepackage{booktabs}
\usepackage{graphicx}
\usepackage{kantlipsum}

\usepackage{booktabs} 
\graphicspath{{figs/}}
\usepackage{braket}

\begin{document}

\title{\huge{Knowledge Distillation in Quantum Neural Network using Approximate Synthesis}}

\author{
\IEEEauthorblockN{Mahabubul~Alam, Satwik~Kundu, Swaroop~Ghosh}

\IEEEauthorblockA{
\textit{School of Electrical Engineering and Computer Science, Penn State University, University Park}\\
\textit{mxa890@psu.edu, satwik@psu.edu, szg212@psu.edu}
}
}

\maketitle

\begin{abstract}

Recent assertions of a potential advantage of Quantum Neural Network (QNN) for specific Machine Learning (ML) tasks have sparked the curiosity of a sizable number of application researchers. 
The parameterized quantum circuit (PQC), a major building block of a QNN, consists of several layers of single-qubit rotations and multi-qubit entanglement operations. The optimum number of PQC layers for a particular ML task is generally unknown. A larger network often provides better performance in noiseless simulations. However, it may perform poorly on hardware compared to a shallower network. Because the amount of noise varies amongst quantum devices, the optimal depth of PQC can vary significantly. Additionally, the gates chosen for the PQC may be suitable for one type of hardware but not for another due to compilation overhead. This makes it difficult to generalize a QNN design to wide range of hardware and noise levels. An alternate approach is to build and train multiple QNN models targeted for each hardware which can be expensive. To circumvent these issues, we introduce the concept of knowledge distillation in QNN using approximate synthesis. 
The proposed approach will create a new QNN network with (i) a reduced number of layers or (ii) a different gate set without having to train it from scratch. Training the new network for a few epochs can compensate for the loss caused by approximation error. Through empirical analysis, we demonstrate $\approx$71.4\% reduction in circuit layers, and still achieve $\approx$16.2\% better accuracy under noise.


\end{abstract}

\maketitle

\section{Introduction}

Quantum computing is advancing rapidly. The community is seeking computational advantages with quantum computers (i.e., quantum supremacy) for practical applications. Recently, Google claimed quantum supremacy with a 53-qubit quantum processor to complete a computational task in 200 seconds that might take 10K years \cite{arute2019quantum} (later rectified to 2.5 days \cite{pednault2019quantum}) on the state-of-the-art supercomputers. Quantum machine learning (QML) is a promising application domain to archive quantum advantage with noisy quantum computers in the near-term. Numerous QML models built upon parametric quantum circuits (PQC), also referred to as quantum neural networks (QNN), are already proposed in the literature \cite{farhi2018classification, killoran2019continuous, schuld2021effect, lloyd2020quantum}. 

\begin{figure*}[]
\vspace{-4mm}
 \begin{center}
    \includegraphics[width=0.9\textwidth]{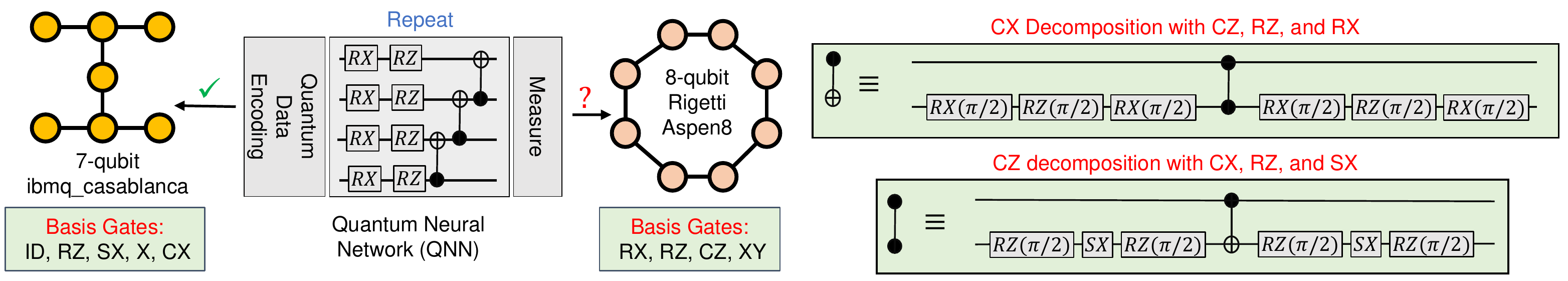}
 \end{center}
 \vspace{-4mm}
\caption{A toy QNN is shown next to two hardware coupling graphs from IBM and Rigetti. They each offer a distinct set of basis gates. A PQC with CX gates may be a good choice for IBM devices. In Rigetti hardware, however, each CX gate is decomposed into one CZ and six additional single qubit gates. This PQC may not be a good fit for the Rigetti devices due to the noise introduced by each gate during computation on actual hardware. Training multiple hardware-specific QNNs from scratch can be costly.
} 
\vspace{-4mm}
\label{fig:demo}
\end{figure*}

A PQC is a quantum circuit with parameterized gates as shown in Fig. \ref{fig:demo}(b). It generally consists of repeated layers of single-qubit rotations (to explore the search space) and multi-qubit operations (to create entanglement). The parameters of PQC can be tuned to attain desired outputs for given inputs (e.g., classifying data samples). QNN models are claimed to be more expressive compared to the classical neural networks \cite{du2020expressive, abbas2021power}. In other words, QNN models have higher capability to approximate a desired functionality compared to the classical models of similar scale (e.g., with same number of tunable parameters/weights). As a result, QNN has a high potential for demonstrating quantum advantage in real applications.

The existing QNNs faces two major adaptability challenges:

\textbf{Noise adaptability:} A large number of PQC layers generally translates to a better QNN performance in noiseless simulation \cite{abbas2021power, farhi2018classification}. However, the near-term quantum devices have a limited number of qubits, and they suffer from various errors such as, decoherence, gate errors, measurement errors and crosstalk. These noises build up quickly as a circuit is scaled up. Therefore, a deep QNN may perform poorly on an actual hardware compared to a shallower network. 
On top of that, the noise levels of a device may vary over time \cite{alam2019addressing}. A trained QNN may perform differently on the same device over time, making it difficult to select an optimal QNN depth for a particular ML task. Training multiple QNN models with different depths and then selecting one based on current target hardware noise levels can be a naive solution to this problem. But, training a QML model from scratch is expensive in terms of time and computational resources \cite{farhi2018classification, killoran2019continuous}.

\textbf{Hardware adaptability:} Different quantum hardware may exhibit varying degrees of noise \cite{ash2019qure}. Furthermore, they may support a different set of basis gates (Fig. \ref{fig:demo}). 
Other gates must be broken down into their component parts. For example, the CX gate which is a native gate in IBM quantum computers will break into 1 CZ gate and 6 other single-qubit gates for Rigetti devices (Fig. \ref{fig:demo}). Similarly, a CZ gate will break down into 1 CX and 6 single-qubit gates for IBM devices. As a result, porting a QNN designed for one piece of hardware (e.g., tailored with CX gates for IBM devices) to another (e.g., Rigetti) is problematic. Creating and training multiple device-specific QNN models is also not a viable option.

A somewhat similar problem exists in the classical domain where an expensively trained (e.g., in a server environment with multiple GPU/CPU) deep neural network (DNN) cannot be run on resource constrained mobile/edge devices. Network pruning, weight sharing, quantization, layer fusion, etc. are some of the techniques available in the classical domain to reduce the size of a pre-trained DNN \cite{neill2020overview}. Knowledge distillation (KD) is another popular DNN compression technique \cite{hinton2015distilling}. In KD, a pre-trained large DNN is used as a guide to train a smaller DNN without sacrificing much performance. 

Inspired from the classical approach, we introduce KD in QNN using approximate synthesis to address the adaptability challenges in QNN. The traditional KD technique uses original data to train the student network with the original labels and pre-trained model outputs guiding the training procedure. However, this can result in a similar amount of training time as learning a new model from scratch. The proposed method avoids such costly training technique by employing approximate synthesis to mimic the pre-trained QNN behavior. Finding an equivalent unitary transformation for an existing quantum circuit using different gates is called approximate synthesis \cite{khatri2019quantum}. Once trained, the PQC of a QNN performs a fixed unitary transformation of the input states. Using approximate synthesis, we can find a comparable PQC that has fewer layers or an entirely different set of gates than the original. Here, the original PQC can act as a guide to create a new network that is more compact and noise-resistant, or that is customized for a completely different hardware. Performance may suffer due to approximation errors which can be recovered by training the new network for a few epochs (far cheaper than training a new QNN from scratch).

\textit{To the best of our knowledge this is the first work to demonstrate approximate synthesis for knowledge distillation in QNN to address adaptability and resilience challenges.}

\section{Preliminaries}

\noindent {\bf{Qubits, Quantum Gates, Measurements \& Quantum Circuit:}} 
Unlike a classical bit, a qubit can be in a superposition state i.e., a combination of $\ket{0}$ and $\ket{1}$ at the same time. A variety of technologies exist to realize qubits such as, superconducting qubits, trapped-ions, to name a few. Quantum gates (e.g., single qubit Pauli-X gate or 2-qubit CNOT gate) modulate the state of qubits and thus perform computations. These gates can perform a fixed computation (e.g., an X gate flips a qubit state) or a computation based on a supplied parameter (e.g. the RY($\theta$) gate rotates the qubit along the Y-axis by $\theta$). A two-qubit gate changes the state of one qubit (\textit{target qubit}) based on the current state of the other qubit (\textit{control qubit}). For example, the CNOT gate flips the target qubit if the control qubit is in $\ket{1}$ state. A quantum circuit contains many gate operations. Qubits are measured to retrieve the final state of a quantum program.

\noindent {\bf{Quantum Noise:}} 
Errors in quantum computing can be broadly classified into, (i) Coherence errors: a qubit can retain its state for a short period (coherence time). The computation needs to be done well within this limit. (ii) Gate errors: quantum gates are realized using microwave/laser pulses. It is impossible to generate and apply these pulses precisely in actual hardware making gate operations erroneous. (iii) Measurement errors: a $\ket{0}$ state qubit can be measured as $\ket{1}$ (or vice versa) due to imprecise measurement apparatus. Execution of multiple gates in parallel can lead to crosstalk errors. Since a large/deep quantum circuit accumulates more errors, a smaller circuit is always preferred for noise-resilience and reliability. 

\noindent {\bf{Quantum Circuit Compilation:}} 
A practical quantum computer generally supports a limited number of single and multi-qubit gates known as \textit{basis gates} or native gates of the hardware. For instance, the current generation of IBM quantum computers have the following basis gates: ID, RZ, SX, X (single-qubit), CNOT (two-qubit). However, the quantum circuit may contain gates that are not native to the target hardware. 
Hence, the gates in a quantum circuit need to be \textit{decomposed} into the basis gates before execution. Besides, the native two-qubit operation may or may not be permitted between all the two-qubit pairs. These limitations in two-qubit operations are also known as \textit{coupling constraints}. Conventional compilers add necessary SWAP gates to meet the coupling constraints. Thus, a compiled circuit depth and gate counts can be significantly higher than the original (compilation overhead). The compilation overhead is dictated by the native gates/connectivity of the target hardware, and the efficiency of the compiler software.

\begin{figure}[b]
\vspace{-4mm}
 \begin{center}
    \includegraphics[width=0.4\textwidth]{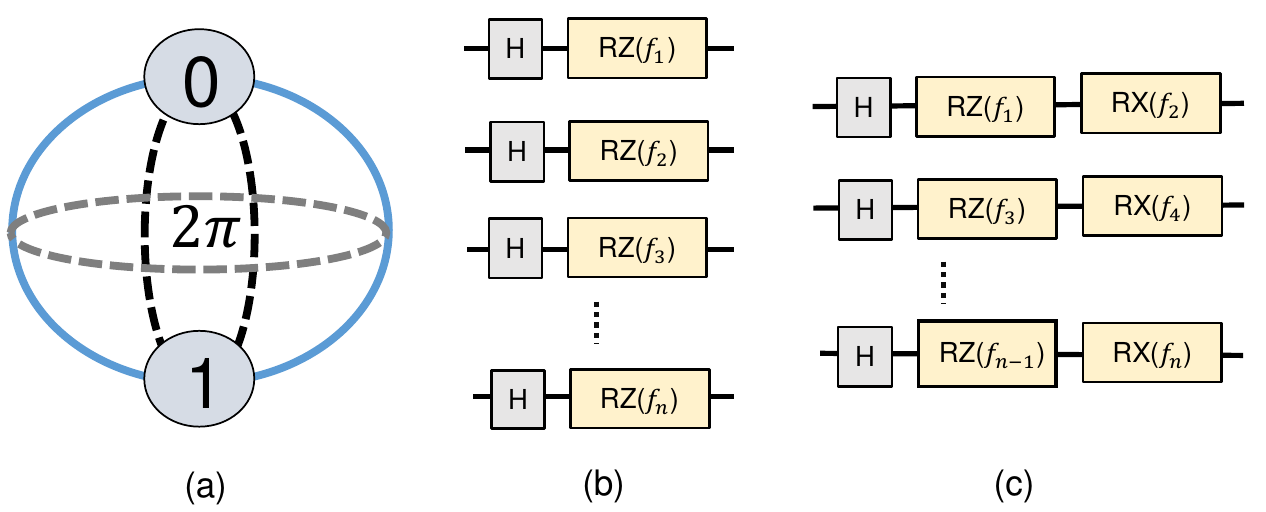}
 \end{center}
 \vspace{-4mm}
\caption{(a) Bloch sphere representation of a qubit. A qubit can be rotated along the X, Y, or Z axis. The states repeat in 2$\pi$ intervals, (b) angle encoding 1:1 (i.e., one continuous variable encoded in a single qubit state), and (c) angle encoding 2:1 (i.e., two continuous variables encoded in a single qubit state).
} 
\vspace{-4mm}
\label{fig:qubit}
\end{figure}

\noindent {\bf{Quantum Neural Network:}} QNN involves parameter optimization of a PQC to obtain a desired input-output relationship. QNN generally consists of three segments: (i) a classical to quantum data encoding or embedding circuit, (ii) a parameterized circuit (PQC), and (iii) measurement operations. A variety of encoding methods are available in the literature \cite{schuld2021effect}. For continuous variables, the most widely used encoding scheme is angle encoding \cite{schuld2021effect, lloyd2020quantum, abbas2021power} where a continuous variable classical feature is encoded as a rotation of a qubit along a desired axis (X/Y/Z). For `n' classical features, we require `n' qubits. For example, RZ(f1) on a qubit in superposition (the Hadamard - H gate is used to put the qubit in superposition) is used to encode a classical feature `f1' in Fig. \ref{fig:qubit}(b). We can also encode multiple continuous variables in a single qubit using sequential rotations (Fig. \ref{fig:qubit}(c)). States produced by a qubit rotation along any axis will repeat in 2$\pi$ intervals (Fig. \ref{fig:qubit}(a)). Therefore, features are generally scaled within 0 to 2$\pi$ in a data pre-processing step. One can restrict the values between -$\pi$ to $\pi$ to accommodate features with both negative and positive values. 

The PQC consists of multiple layers of entangling operations and parameterized single-qubit rotations. The entanglement operations are a set of multi-qubit operations between the qubits to generate correlated states \cite{lloyd2020quantum}. The following parametric single-qubit operations search through the solution space. This combination of entangling and rotation operations is referred to as a parametric layer (PL) in the literature. The optimal number of PL for any given ML task is generally unknown. The problem is similar to choosing the number of hidden layers/neurons in a classical DNN. In practice, one needs to go through multiple training iterations with different number of PL's to come up with a compact network. A compact network can offer better noise-resilience/reliability, and lower latency/faster execution during inference. There is a wide variety of choices available for PL. The work in \cite{sim2019expressibility} analyzed 19 widely used PL architectures from literature. We also use these 19 architectures to evaluate our proposed methodologies (cX refers to circuit X in \cite{sim2019expressibility}). Due to the differences in compilation overhead, a PL that works well for one hardware platform may not be ideal for another.

A widely used PL is shown in Fig. \ref{fig:demo}. Here, CNOT gates between neighboring qubits create the entanglement, and rotations along X \& Z-axis using RX($\theta$) \& RZ($\theta$) operations define the search space.

\noindent {\bf{Knowledge Distillation:}}
This is a model compression technique \cite{hinton2015distilling} where a small (student) model is trained to match a large (teacher) pre-trained model. The teacher model's knowledge is transferred to the student by minimizing a loss function that aims to match both softened teacher logits and ground-truth labels. The logits are softened by using a "temperature" scaling function in the softmax, effectively smoothing out the probability distribution and revealing the teacher-learned inter-class relationships.

\noindent {\bf{Approximate Synthesis:}}
The intended state transformation of a quantum gate can be represented by a unitary matrix. An n-qubit quantum circuit can be represented by a unitary matrix $U$ of size $2^nx2^n$ (tensor product of the gate matrices). Quantum circuit synthesis is the process of finding a new circuit with a different set of gates whose unitary matrix representation ($V$) is exact (exact synthesis) or approximate (approximate synthesis) equivalent of the original ($U$). A synthesis algorithm tries to minimize the distance between $U$ and $V$ ($||U-V||$) using metrics such as L1 norm, L2 norm, diamond norm and frobenius norm \cite{davis2020towards, khatri2019quantum, kliuchnikov2014asymptotically}. L1 norm is the sum of the absolute element-wise differences ($\sum_{i=1}^{2^n} \sum_{j=1}^{2^n} |U_{ij} - V_{ij}|$). L2 norm is the euclidean distance between the matrices ($\sqrt{\sum_{i=1}^{2^n} \sum_{j=1}^{2^n} (U_{ij} - V_{ij})^2}$). 
If $V$ is the unitary matrix representation of a PQC, the parameters can be updated by a conventional optimizer (e.g., Nelder-Mead, BFGS, dual\_annealing, etc. 
to minimize $||U-V||$. Note that, it is not guaranteed that the optimizer will converge to an exact solution (distance = 0) for any given $U$ and the choice of PQC for $V$. More often that not, it will generate a $V$ that is a close approximation of $U$. The outcome will largely depend on the choice of the PQC for $V$, dimension of $U$, chosen optimization method (local/global), and the computing resource/time allocated for optimization.  

\section{Related Works}

The work in \cite{khatri2019quantum} introduced approximate synthesis for quantum circuit compression/depth reduction, arbitrary quantum state preparation, and noise reduction. This work showed that quantum circuits with high depth/gate-count but small number of qubits (two/three) can often be approximated with a smaller circuit (lesser gates/depth). Because quantum gates are erroneous, the benefits of circuit size reduction can outweigh the approximation error from such compression. 
However, the approximation error becomes too large when the number of qubit is increased. 

Recent works have used the divide and conquer approach to apply approximate synthesis on larger circuits \cite{davis2020towards, wu2020qgo}. Here, a large quantum circuit is divided into smaller sub-circuits (2-4 qubits). Smaller representation of this sub-circuits are found through approximate synthesis. Later, these smaller representations replace the original sub-circuits \cite{wu2020qgo}. It's worth noting that most previous works have used approximate synthesis to improve arithmetic quantum circuits or state-preparation circuits. These circuits have a very small tolerable margin of error. 
Since QNN or any other ML model approximates a function (e.g., classification/regression), the margin of error in synthesis can be higher. 
In fact, it may provide better performance on hardware if the benefits with reduction of gate count/depth outweighs the loss incurred due to approximation. However, there is a lack of research on approximation synthesis of QNN circuits.


\section{Proposed Methodology}

\begin{figure}[]
\vspace{-4mm}
 \begin{center}
    \includegraphics[width=0.4\textwidth]{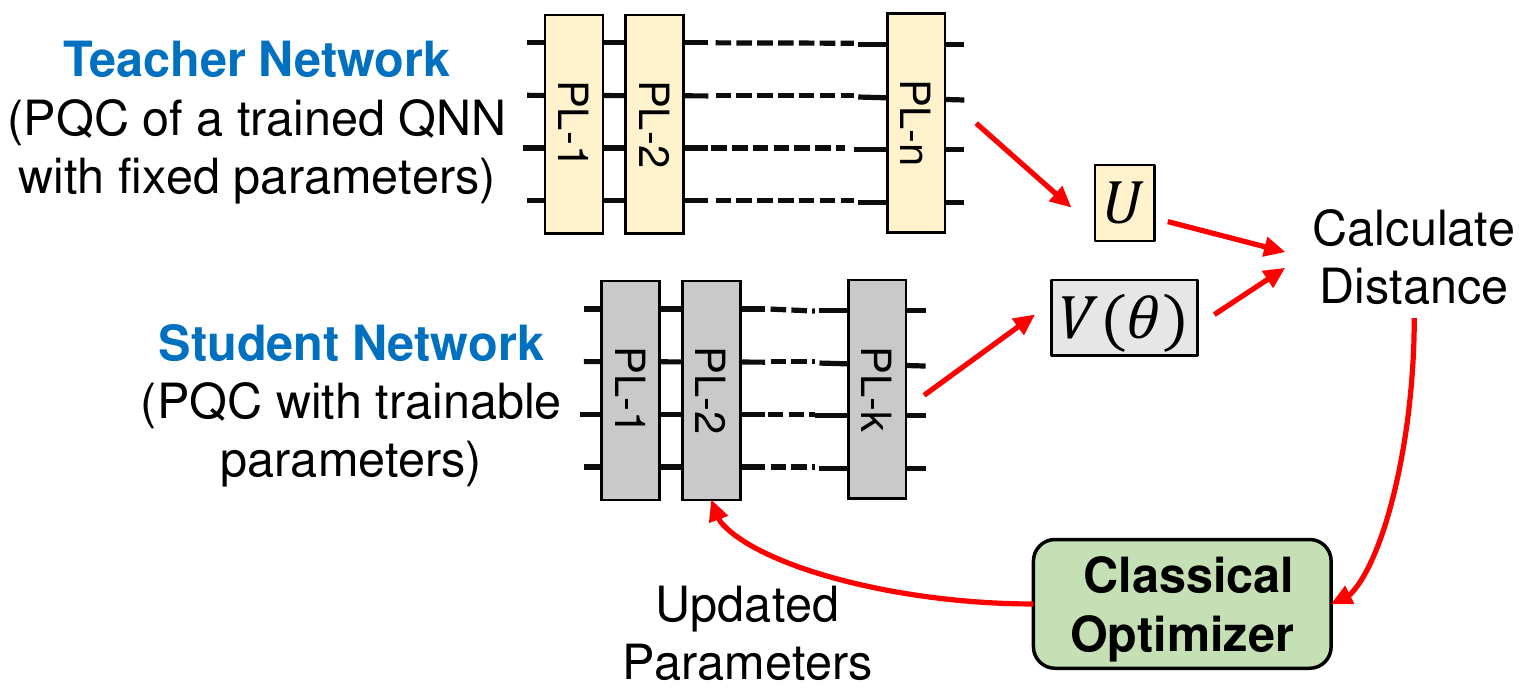}
 \end{center}
 \vspace{-4mm}
\caption{Proposed teacher-student knowledge distillation framework for QNN where a student (PQC with trainable parameters) is guided by a teacher (PQC of a trained QNN with fixed set of parameters) to learn the teachers knowledge (i.e., mimic its unitary transformation).
} 
\vspace{-4mm}
\label{fig:teacherstudent}
\end{figure}

We follow a teacher-student knowledge distillation framework (Fig. \ref{fig:teacherstudent}) \cite{hinton2015distilling} where the teacher is the PQC of a pre-trained QNN. The teacher PQC's unitary matrix representation is the knowledge that is learnt during the training. The student learns to mimic the unitary transformation of the teacher PQC through approximate synthesis without being trained on the original dataset. Later, the loss in performance due to approximation error is mitigated through training the student network for a small number of epochs.

\begin{figure}[b]
\vspace{-4mm}
 \begin{center}
    \includegraphics[width=0.42\textwidth]{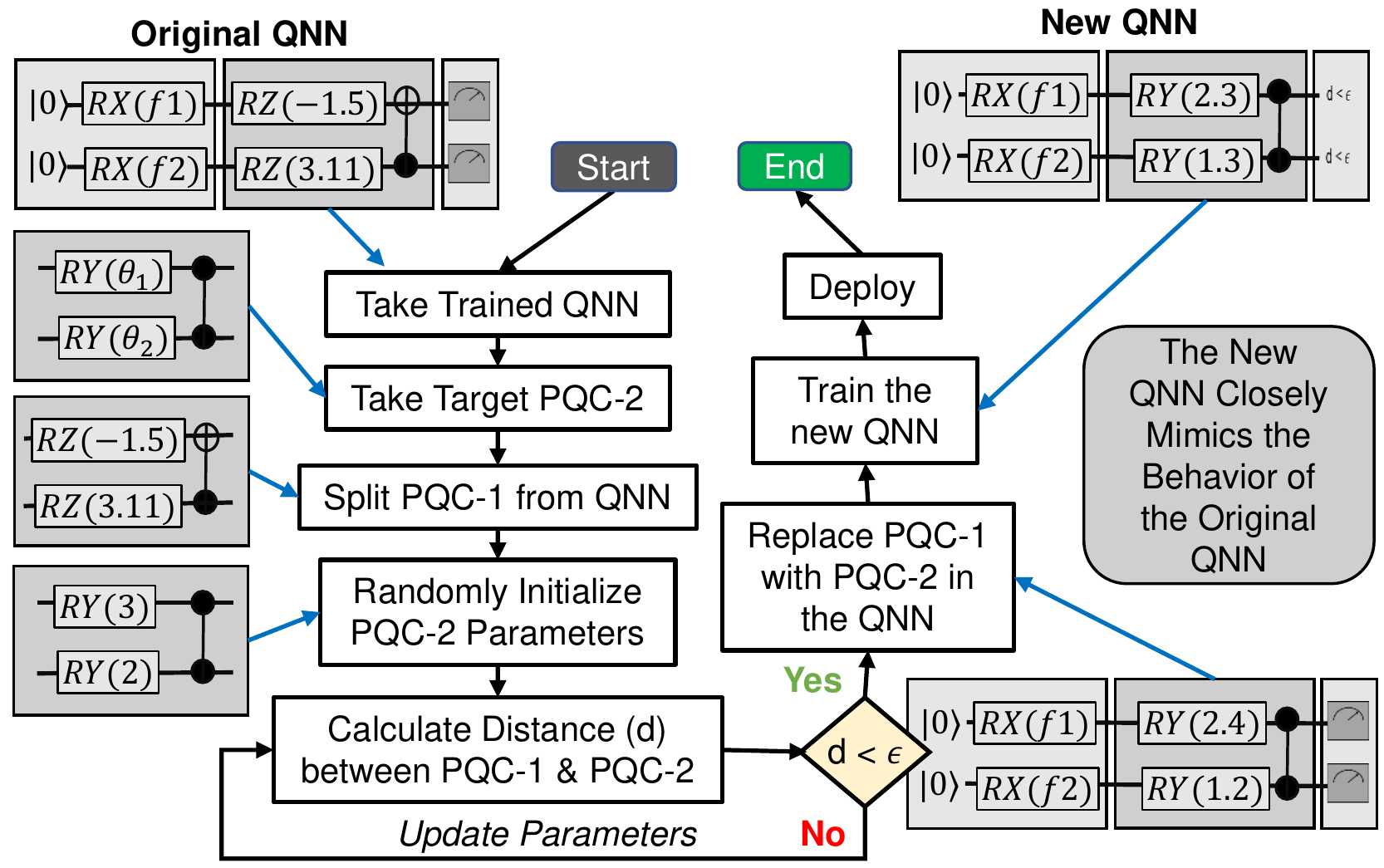}
 \end{center}
 \vspace{-4mm}
\caption{Transferring knowledge from a pre-trained QNN to a new QNN using approximate synthesis. Approximation error may affect the performance which is partially recovered by training the student network for a few epochs.
} 
\vspace{-4mm}
\label{fig:flow}
\end{figure}

{\bf{Overview:}}
The knowledge distillation process (Fig. \ref{fig:flow}) starts with a pre-trained QNN and a desired PQC architecture (PQC-2). The pre-trained network can be a large/short-depth QNN with its basic building blocks - a quantum-to-classical data encoding circuit, a PQC with fixed/trained parameters (PQC-1), and measurements. PQC-1 is separated from the pre-trained QNN. We also freeze its parameters. Note that both these PQC's have the same number of qubits (n) and can be represented by unitary matrices of dimension $2^nx2^n$. Since, PQC-1 parameters are frozen, its matrix representation is a fixed unitary. On the other hand, PQC-2 parameters are trainable. Therefore, its corresponding unitary transformation is a function of its parameters. For any random set of values of PQC-2 parameters, the distance between these two unitary matrices can be large. This distance can be calculated as a single scaler quantity using various types of norms, e.g., L1, L2, etc. 

After choosing a suitable norm as the distance metric, we can use a classical optimizer to update the PQC-2 parameters and minimize this distance. If the optimization procedure ends with a sufficiently low value of this distance, then we have a new PQC (PQC-2 with optimized parameters) that is a close approximation of PQC-1. After plugging in PQC-2 in place of PQC-1, we get a new QNN that performs approximately similar to the original QNN. If the distance is not sufficiently small, the new QNN performance can be inferior to the original QNN. To recuperate this loss, we can train the new QNN with the original dataset for a few epochs.

{\bf{Distance Metric:}}
We use following metric (based on the Hilbert-Schmidt inner product) as cost function for approximate synthesis: $d = 1 - \frac{Tr(U^{\dagger}V)}{dim(U)}$ where U and V are the unitary matrix representations of PQC-1 and PQC-2. The inverse of a unitary matrix is its conjugate transpose. Therefore, if the synthesis succeeds and $U$ is close to $V$, the product $U^{\dagger}V$ = $I_N$ where $I$ is the identity matrix and $N$ = dim($U$). Furthermore, the maximum magnitude that the trace of a unitary matrix can have is its size $N$, which occurs at the identity (up to a phase). The closer $U^{\dagger}V$ is to identity, the closer $Tr(U^{\dagger}V)$ is to N, and thus, the distance (d) closer to 0. Note, existing works on approximate synthesis have also used this metric \cite{khatri2019quantum, davis2020towards} because (i) it has computational advantage over other methods, (ii) the distance becomes 0 if and only if the matrices match exactly, (iii) it scales well with the size of the problem, and (iv) it has operational meaning.


\begin{figure*}[htb]
\vspace{-4mm}
 \begin{center}
    \includegraphics[width=0.86\textwidth]{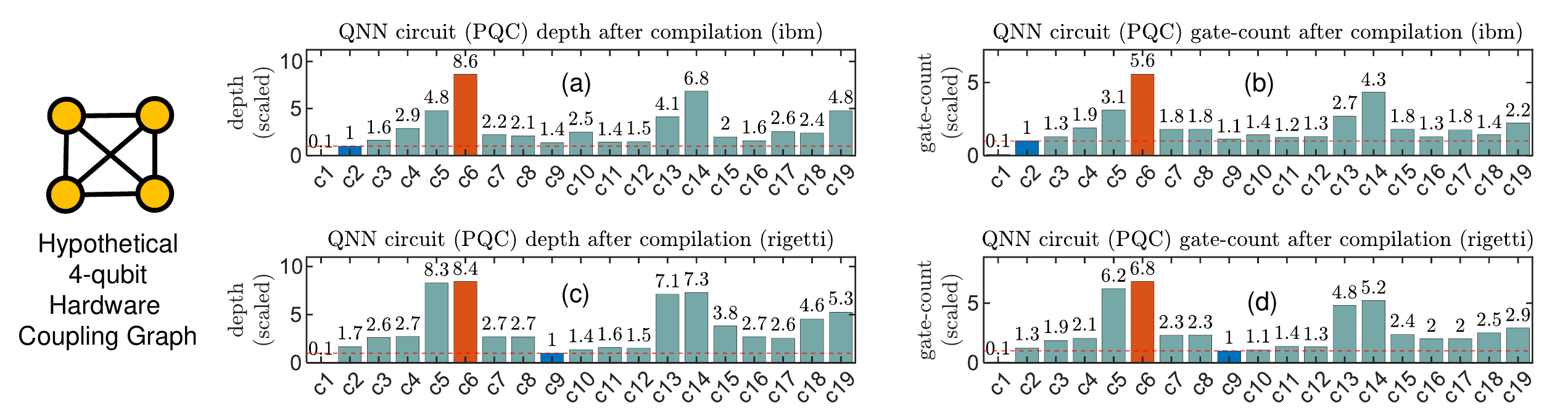}
 \end{center}
 \vspace{-4mm}
\caption{Depth and gate-count of the 19 single-layer 4-qubit PQC's in \cite{sim2019expressibility} after compilation with a hypothetical 4-qubit hardware with 3 different gate sets (IBM \& Rigetti). Here, cX refers to circuit X in \cite{sim2019expressibility}.
}
\vspace{-4mm}
\label{fig:compilation}
\end{figure*}

\begin{figure*}[hbt]
\vspace{-1mm}
 \begin{center}
    \includegraphics[width=0.9\textwidth]{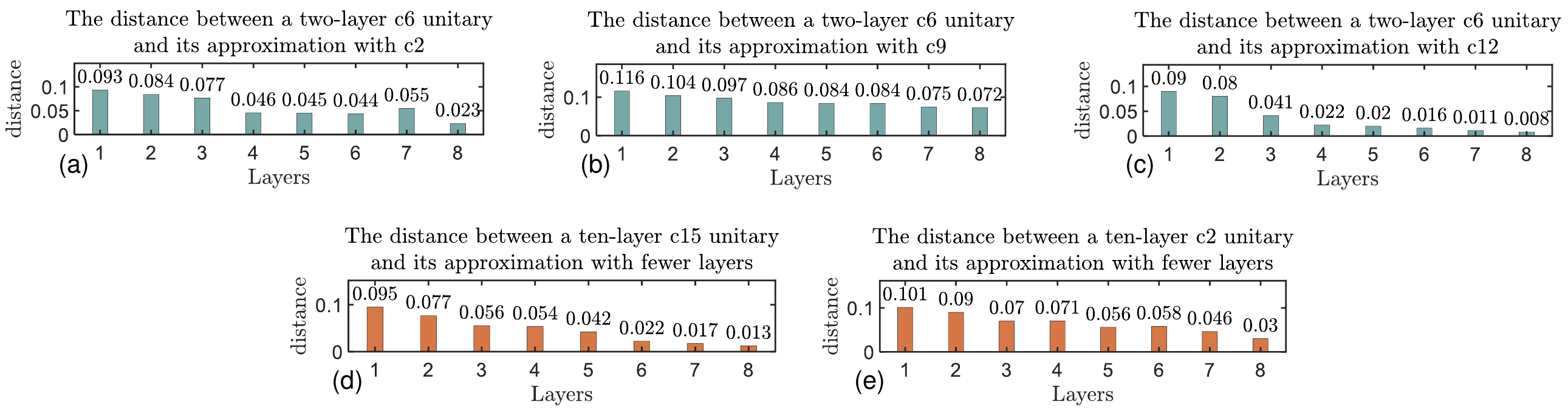}
 \end{center}
\vspace{-4mm}
\caption{In (a), (b), and (c), we show the distance between a 4-qubit 2-layer c6 PQC, and its compressed versions with 1-8 layer c2, c9, and c12 PQC's. In (d), we show the distance between a 10-layer c15 PQC and its compressed versions. (e) shows a similar analysis with c2. In all cases, the base PQC is trained to classify the Iris dataset.
}
\vspace{-4mm}
\label{fig:approximation}
\end{figure*}

{\bf{Distance Minimization:}}
Any classical optimizer can be used to minimize the distance. We can choose any gradient-based (e.g., Adam, AdaGrad, SGD, BFGS, etc.) or gradient-free (e.g., Nelder-Mead) optimizer for this task. 
These algorithms, however, execute local searches and are prone to get stuck in local minima. To increase the chances of getting closer to a global minima, one can run multiple procedures with different random seeds with these local optimizers. Alternatively, one can use global optimization methods, e.g., differential evolution, simulated annealing, etc., for better chances of getting an optimal solution at the cost of higher runtime. We use the dual annealing optimizer from the SciPy library which is an extension of the classic simulated annealing. 

{\bf{Choice of PQC:}}
Approximate synthesis can be used to (i) compress a QNN (same PL architecture, lower number of layers), and (ii) find an efficient QNN for a target hardware platform (PL tailored for the target hardware). Therefore, the choice of PQC depends on the user objective. A PQC may be an excellent choice for one hardware platform but unsuitable for another. To further illustrate this issue, we have used two basis gate-sets (IBM \& Rigetti), and compiled a total of 19 benchmark PQC circuits from \cite{sim2019expressibility} with the qiskit compiler 
for a hypothetical 4-qubit fully-connected hardware. Fig \ref{fig:compilation} shows the relative depth and gate counts with identical number of layers. Note that, c2 has the lowest-depth and gate-count with IBM gate-set as shown in Fig. \ref{fig:compilation}(a)\&(b) (we omit c1 from comparison since it does not have any entanglement). The c6 PQC can be 8.6X deeper and can have 5.6X larger gate count due to compilation overhead. Note that c2 PQC uses CX operation between the neighboring qubits to create entanglement which is native to the IBM gate-set, and therefore, more suitable for this gate-set. Similarly, c9 is suitable for Rigetti gate-set since it uses CZ for entanglement which is native to this gate-set (Fig. \ref{fig:compilation}(c)\&(d)). The c2 PQC with a similar number of PL can be 1.7X deeper, and the gate count can be 1.3X higher compared to c9 on Rigetti gate-set. As demonstrated by these results, a QNN optimized for one type of hardware may perform poorly on another if executed naively following compilation due to compilation overhead.

{\bf{Approximation Error:}}
Approximate synthesis may not always produce an exact solution. The outcome will depend on the choice of PQC, the size of the PQC's in terms of qubits, the chosen optimization procedure, and the computational resource/time allocated for optimization. A very shallow PQC-2 may not have sufficient expressive power to mimic the unitary transformation of PQC-1 \cite{sim2019expressibility}. It is noted that approximate synthesis does not scale well with qubit-size \cite{khatri2019quantum, davis2020towards} due to significant approximation error. Moreover, it may search through a local minima. If a global optimizer is terminated after a certain number of function evaluations, it may not be able to reach an optimal solution. Therefore, approximate synthesis may incur approximation error and compromise the QNN performance.

{\bf{Training the Student Network:}}
Performance loss due to approximation error can be mitigated by training the student model. The compressed model parameters can be used as the initial seed. If the approximation error is minor, a small number of training epochs with the original dataset may be sufficient to recover the majority of the loss else a large number of epochs might be necessary. Nonetheless, the process can be significantly less expensive than training a new model from scratch using random seeds.

\section{Evaluation}

\subsection{Setup}
\indent \indent {\bf{Framework:}}
We use the PennyLane and TensorFlow packages to train the QNN models. Qiskit is used for circuit compilations and noise simulations. SciPy package is used for optimization. All numerical experiments are run on a Intel Core-I7-10875H CPU with 16GB of RAM. The teacher models are trained for 10 epochs and the student models are trained for 2 epochs using the Adam optimizer (learning rate = 0.2, beta\_1 = 0.9, beta\_2 = 0.999, epsilon = $e^{-07}$). We use the dual\_annealing optimizer as the classical optimizer in approximate synthesis (initial\_temp = 5230.0, restart\_temp\_ratio = $2e^{-05}$, visit = 2.62, accept = -5.0). The parameter values are bounded between -$\pi$ to $\pi$ and the maximum number of function calls is restricted to 1000.

{\bf{Datasets:}} We use the Iris, MNIST, and Fashion-MNIST datasets. The Iris dataset has 150 samples of 3 different classes (50/class). Each sample has 4 features. The MNIST and Fashion-MNIST datasets both have 60000 training samples and 10000 test samples that belong to 10 different classes. Each sample has 28x28 features. To reduce simulation time, we have created 6 smaller 3-class datasets from the MNIST and Fashion-MNIST datasets - MNIST179, MNIST246, MNIST358, FASHION012, FASHION345, and FASHION678. First, we reduce the dimension of MNIST and Fashion-MNIST datasets to 8 using a convolutional auto-encoder \cite{alam2021quantum}. Later, we pick 750 samples of 3 different classes (250/class) to create a smaller dataset (e.g., 750 samples are drawn from classes 0, 1, and 2 for FASHION012). We divide this 750 samples to a train (600) and validation set (150).

{\bf{QNN Networks, Qubits, Encoding \& Loss Function:}}
We use 4-qubit QNN models in all the QNN's. The 19 PQC architectures from \cite{sim2019expressibility} are used in our work. We use 1 variables/qubit encoding (Fig. \ref{fig:qubit}(b)) for Iris, and 2 variables/qubit encoding (Fig. \ref{fig:qubit}(c)) for MNIST and Fashion-MNIST. Pauli-Z expectation values of the qubits are taken as the QNN outputs. We feed these outputs to a fully-connected classical dense layer with 3 neurons to facilitate 3-class classification problems \cite{alam2021quantum}. A Softmax layer transforms the output of this neurons to class probabilities. We use the categorical cross-entropy loss. 

{\bf{Metrics:}} We prefer to evaluate teacher models, student models, and trained student models using classification accuracy on the training and validation sets. 

\subsection{Results}

\indent \indent {\bf{QNN Retargeting:}}
Approximate synthesis can be used to find an equivalent QNN to a pre-trained QNN that is more suitable for a target quantum device. The pre-trained QNN might have been trained for a different device. The PL of the new PQC can be chosen based on the basis gates of the new hardware. For example, the c6 PQC consists of several CRX gates which are not native to IBM and Rigetti hardware. A single-layer c6 PQC has 8.6X, and 8.4X higher number of gates than a single-layer c2, and c9 PQC when compiled with the IBM and Rigetti gate-sets, respectively. If a pre-trained QNN uses c6 PQC, it can be approximated with c2 and c9 PQC's for IBM and Rigetti devices. We show the distance between a 2-layer c6 4-qubit PQC trained to classify the Iris dataset and its c2 approximation with 1-8 layers in Fig. \ref{fig:approximation}(a). Note that, with a single-layer, the distance is quite large (0.093) which decreases consistently with added layers (0.023 with 8-layers). Also note that an 8-layer c2 PQC has less than half the gates of a 2-layer c6 PQC when compiled with the IBM gate-set. We may get to an even lower distance (closer to exact synthesis) by increasing the c2 layer and still maintain the number of gates lower than 2-layer c6 PQC. Note that, with every PQC, we may not explore similar spaces in the Hilbert-space \cite{sim2019expressibility}. Therefore, a PQC may perform better over another to approximate the target PQC. For instance, with 8-layers, the c12 PQC gets much closer to the 2-layer c6 PQC compared to c9 as shown in Fig. \ref{fig:approximation}(b)\&(c) (0.008 vs. 0.072). Therefore, c12 can be a better choice to approximate c9 for Rigetti devices, both having considerably lower gate count/depth upon compilation with Rigetti gate-set (Fig. \ref{fig:compilation}). 

{\bf{QNN Compression:}}
Since noise increases with the circuit size, compression can be extremely useful to mitigate the impact of noise during inference of the model on an actual hardware. For example, we take 10-layer c15, and c2 PQC's and trained them to classify the Iris dataset. Later, we approximate these PQC's with 1-8 layers. The results are shown in Fig. \ref{fig:approximation}(d)\&(e). The distances consistently decrease with added layers (0.095 with 1-layer to 0.013 with 8-layers for c15). Note that, reducing a single layer can be extremely significant because it can reduce, (i) noise accumulation, and (ii) the latency or increase the speed during model inference. For instance, an 8-layer c15 will require 80\% of the execution time compared to a 10-layer c15 PQC. That will translate to reduced coherence errors. On top of that, it will have 20\% less gates that will result in lesser accumulation of gate errors.

{\bf{Performance after Approximation and Training:}}
The approximated model 
performances can be inferior in noiseless simulation. To illustrate this, we trained the MNIST and Fashion-MNIST datasets with 2-layer C6 PQC's and approximated them with 6-layer c2 and c9 PQC's (QNN Retargeting). Similarly, we have also trained a 7-layer c15 PQC and later approximated it with 2-5 layers c15 PQC's (QNN Compression). The results are tabulated in Table \ref{tab:1} and \ref{tab:2}. For some datasets (e.g., FASHION345), the approximated models provide similar performance compared to the base model (0.91 and 0.925 accuracy over the training set with 6-layer c2 and c9 compared to 0.9316 with 2-layer c6). However, in some cases (e.g., MNIST179), the performance is significantly compromised (0.5550 and 0.6899 accuracy over the training set with 6-layer c2 and c9 compared to 0.9033 with 2-layer c6). As expected, the performance loss is higher with more compression (Table \ref{tab:2}). For example, with 4-layers, the accuracy on MNIST179 training set is 0.4933 which increases to 0.6499 with 5-layers. However, the compressed models performance remains inferior to the 7-layer c15 model for all the datasets. After training the retargeted/compressed models only for two epochs (with approximated model parameters as the initial seeds), we get closer to the base model as evident from Table \ref{tab:1} and \ref{tab:2}.

\begin{figure}[]
 \begin{center}
    \includegraphics[width=0.35\textwidth]{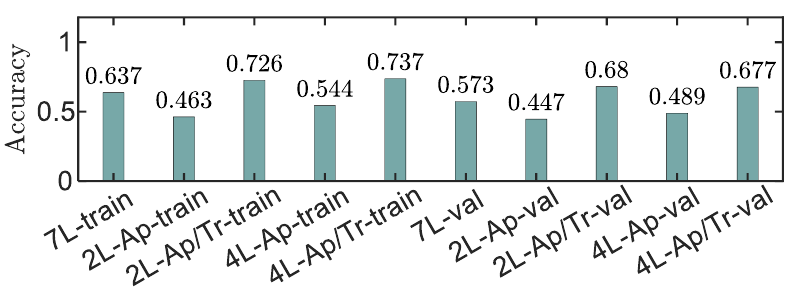}
 \end{center}
 \vspace{-4mm}
\caption{Measured average accuracy of the 7L c15 base models, 2L/4L approximated (Ap), and approximated \& trained models (Ap/Tr) across all the training and validation datasets on two hardware emulators - FakeMelbourne, and FakeAlmaden available in the Qiskit framework.
}
\vspace{-4mm}
\label{fig:noise}
\end{figure}

{\bf{Performance under Noise:}}
To study the performance after approximation, we pick all the 7-layer c15 base models, 2/4 layer approximated models (Ap), and the approximated and trained models (Ap/Tr) from Table \ref{tab:2} and measured their accuracy on two hardware emulators available in Qiskit - FakeMelbourne and FakeAlmaden. We used emulators rather than actual hardware because (i) access to actual hardware is still extremely limited and (ii) generating statistically significant results would require an exorbitantly lengthy period of access to hardware. These hardware emulators model the gate errors (using depolarizing noise channel), decoherence errors (using thermal relaxation), and measurement errors (using bit-flipping probabilities) of the devices. The noise parameters are obtained from actual hardware calibration. FakeMelbourne has the following noise attributes - avg. single/two-qubit gate error: 0.104\%/3.14\%, avg. T1: 56.07 ${\mu}s$, avg. T2: 55.5 ${\mu}s$, avg. single/two-qubit gate-execution time: 68.57$ns$/902.9$ns$, and avg. measurement error: 5.63\%. FakeAlmaden - avg. single/two-qubit gate error: 0.09\%/2.38\%, avg. T1: 86.78 ${\mu}s$, avg. T2: 64.31 ${\mu}s$, avg. single/two-qubit gate-execution time: 35.55$ns$/405.8$ns$, and avg. measurement error: 5.35\%. We show the average of the results over all the datasets in Fig. \ref{fig:noise}. Note that, for both the training and validation sets, the approximated and trained student models provide significantly higher accuracy compared to the base model. For example, we get 0.726/0.737 (0.68/0.677) average accuracy with 2L/4L models compared to 0.637 (0.573) of the base 7L model on the training (validation) sets.

\begin{table}[]
\centering
\vspace{-4mm}
\caption{Accuracy of the trained models with two-layer c6 PQC, their approximated counterparts with six-layer c2/c12 PQC's (Ap), and the approximated \& trained models (Ap/Tr)}
\vspace{-1em}
\label{tab:1}
\scalebox{0.75}{
\begin{tabular}{@{}|c|c|c|c|c|c|c|@{}}
\toprule
\multirow{2}{*}{Data-set} & \multirow{2}{*}{Subset} & \multicolumn{5}{c|}{Accuracy} \\ \cmidrule(l){3-7} 
 &  & \begin{tabular}[c]{@{}c@{}}C6\\   (2L)\end{tabular} & \begin{tabular}[c]{@{}c@{}}C2\\   Ap (6L)\end{tabular} & \begin{tabular}[c]{@{}c@{}}C2\\   Ap/Tr (6L)\end{tabular} & \begin{tabular}[c]{@{}c@{}}C9\\   Ap (6L)\end{tabular} & \begin{tabular}[c]{@{}c@{}}C9\\   Ap/Tr (6L)\end{tabular} \\ \midrule
\multirow{2}{*}{MNIST179} & Train & 0.9033 & 0.5550 & 0.8666 & 0.6899 & 0.8000 \\ \cmidrule(l){2-7} 
 & Val & 0.8456 & 0.5436 & 0.8590 & 0.6308 & 0.7785 \\ \midrule
\multirow{2}{*}{MNIST246} & Train & 0.9599 & 0.7216 & 0.8766 & 0.9133 & 0.9116 \\ \cmidrule(l){2-7} 
 & Val & 0.9333 & 0.7400 & 0.7666 & 0.8933 & 0.9333 \\ \midrule
\multirow{2}{*}{MNIST358} & Train & 0.9300 & 0.8233 & 0.9333 & 0.8550 & 0.9166 \\ \cmidrule(l){2-7} 
 & Val & 0.8733 & 0.7933 & 0.8999 & 0.8600 & 0.8733 \\ \midrule
\multirow{2}{*}{FASHION012} & Train & 0.9533 & 0.7200 & 0.9116 & 0.8949 & 0.9333 \\ \cmidrule(l){2-7} 
 & Val & 0.9133 & 0.6866 & 0.9133 & 0.8666 & 0.8733 \\ \midrule
\multirow{2}{*}{FASHION345} & Train & 0.9316 & 0.9100 & 0.9333 & 0.9250 & 0.9499 \\ \cmidrule(l){2-7} 
 & Val & 0.8866 & 0.8600 & 0.9399 & 0.8799 & 0.9133 \\ \midrule
\multirow{2}{*}{FASHION678} & Train & 0.9800 & 0.8283 & 0.9566 & 0.9283 & 0.9666 \\ \cmidrule(l){2-7} 
 & Val & 0.9266 & 0.7533 & 0.8799 & 0.8799 & 0.9266 \\ \bottomrule
\end{tabular}%
}
\vspace{-1em}
\end{table}

\section{Discussion}
Approximate synthesis does not scale well with qubit-size \cite{khatri2019quantum, kliuchnikov2014asymptotically} i.e., the approximation error can grow with qubit size. To illustrate this issue, we took 200 instances of 2, 3, 4, 5, and 6-qubit \& 6-layer c2 PQC with random parameters and approximated them with 4-layer PQC's (40/qubit-size). Later, we measured the fidelity between the states prepared by the 6-layer PQC's and their 4-layer approximations. Fidelity is a widely used to measure the distance between two quantum states (Fidelity($\ket{\psi_a}$, $\ket{\psi_b}$) = $|\langle\psi_a|\psi_b\rangle|^2$, and $\ket{\psi_a} = U_a\ket{0}$). 


We obtain very low approximation error with 2/3 qubit PQCs, as indicated by the higher average fidelity scores of 0.999 and 0.949, respectively. Additionally, the average fidelity scores for 4/5 qubit PQCs are moderate at 0.796 and 0.775. However, 6-qubit PQCs show poor average fidelity score of 0.497. The results indicate that the vanilla approximate synthesis approach can be very useful for QNNs that have higher depth (many PQC layers) but a smaller number of qubits. Many NISQ-era QNN models sacrifice depth for smaller number of qubits \cite{perez2020data, lloyd2020quantum}. These models encode high-dimensional classical data into a small number of qubits using sequential rotations. Once the data is loaded, a PQC transforms it to a output that is used for the intended ML task. Approximate synthesis can be extremely beneficial for such QNNs. The vanilla approach may be less useful for QNNs with a large number of qubits due to the large approximation error. However, we can always choose the divide and conquer approach in such cases \cite{wu2020qgo, khatri2019quantum} where the PQC can be divided into smaller (2/3-qubit) blocks with large depths, and approximated with shorter depth. This is the topic of further explorations.

\begin{table}[]
\vspace{-4mm}
\caption{Accuracy of the trained models with seven-layer c15 PQC, its approximated versions with 2, 4, \& 5 layer PQC's (Ap), and the approximated \& trained models (Ap/Tr)}
\vspace{-1em}
\label{tab:2}
\scalebox{0.75}{
\begin{tabular}{@{}|c|c|ccccccc|@{}}
\toprule
\multirow{2}{*}{Data-set} & \multirow{2}{*}{Subset} & \multicolumn{7}{c|}{Accuracy (c15)} \\ \cmidrule(l){3-9} 
 &  & \multicolumn{1}{c|}{7L} & \multicolumn{1}{c|}{\begin{tabular}[c]{@{}c@{}}2L\\ Ap\end{tabular}} & \multicolumn{1}{c|}{\begin{tabular}[c]{@{}c@{}}2L\\ Ap/Tr\end{tabular}} & \multicolumn{1}{c|}{\begin{tabular}[c]{@{}c@{}}4L\\ Ap\end{tabular}} & \multicolumn{1}{c|}{\begin{tabular}[c]{@{}c@{}}4L\\ Ap/Tr\end{tabular}} & \multicolumn{1}{c|}{\begin{tabular}[c]{@{}c@{}}5L\\ Ap\end{tabular}} & \begin{tabular}[c]{@{}c@{}}5L\\ Ap/Tr\end{tabular} \\ \midrule
\multirow{2}{*}{MNIST179} & Train & \multicolumn{1}{c|}{0.8333} & \multicolumn{1}{c|}{0.4099} & \multicolumn{1}{c|}{0.7749} & \multicolumn{1}{c|}{0.4933} & \multicolumn{1}{c|}{0.7816} & \multicolumn{1}{c|}{0.6499} & 0.7583 \\ \cmidrule(l){2-9} 
 & Val & \multicolumn{1}{c|}{0.8120} & \multicolumn{1}{c|}{0.3557} & \multicolumn{1}{c|}{0.7718} & \multicolumn{1}{c|}{0.4697} & \multicolumn{1}{c|}{0.7248} & \multicolumn{1}{c|}{0.6375} & 0.7046 \\ \midrule
\multirow{2}{*}{MNIST246} & Train & \multicolumn{1}{c|}{0.8383} & \multicolumn{1}{c|}{0.5249} & \multicolumn{1}{c|}{0.7433} & \multicolumn{1}{c|}{0.5483} & \multicolumn{1}{c|}{0.8483} & \multicolumn{1}{c|}{0.4416} & 0.8166 \\ \cmidrule(l){2-9} 
 & Val & \multicolumn{1}{c|}{0.7799} & \multicolumn{1}{c|}{0.5000} & \multicolumn{1}{c|}{0.6466} & \multicolumn{1}{c|}{0.4466} & \multicolumn{1}{c|}{0.8466} & \multicolumn{1}{c|}{0.3866} & 0.7400 \\ \midrule
\multirow{2}{*}{MNIST358} & Train & \multicolumn{1}{c|}{0.8033} & \multicolumn{1}{c|}{0.5083} & \multicolumn{1}{c|}{0.7149} & \multicolumn{1}{c|}{0.6133} & \multicolumn{1}{c|}{0.8149} & \multicolumn{1}{c|}{0.6549} & 0.7799 \\ \cmidrule(l){2-9} 
 & Val & \multicolumn{1}{c|}{0.5933} & \multicolumn{1}{c|}{0.5000} & \multicolumn{1}{c|}{0.6333} & \multicolumn{1}{c|}{0.5799} & \multicolumn{1}{c|}{0.6933} & \multicolumn{1}{c|}{0.4733} & 0.6533 \\ \midrule
\multirow{2}{*}{FASHION012} & Train & \multicolumn{1}{c|}{0.9083} & \multicolumn{1}{c|}{0.5683} & \multicolumn{1}{c|}{0.8433} & \multicolumn{1}{c|}{0.7383} & \multicolumn{1}{c|}{0.8849} & \multicolumn{1}{c|}{0.8333} & 0.8866 \\ \cmidrule(l){2-9} 
 & Val & \multicolumn{1}{c|}{0.8999} & \multicolumn{1}{c|}{0.5799} & \multicolumn{1}{c|}{0.7466} & \multicolumn{1}{c|}{0.7266} & \multicolumn{1}{c|}{0.8199} & \multicolumn{1}{c|}{0.8600} & 0.8666 \\ \midrule
\multirow{2}{*}{FASHION345} & Train & \multicolumn{1}{c|}{0.8966} & \multicolumn{1}{c|}{0.4383} & \multicolumn{1}{c|}{0.8833} & \multicolumn{1}{c|}{0.6250} & \multicolumn{1}{c|}{0.9033} & \multicolumn{1}{c|}{0.8316} & 0.9083 \\ \cmidrule(l){2-9} 
 & Val & \multicolumn{1}{c|}{0.8666} & \multicolumn{1}{c|}{0.4600} & \multicolumn{1}{c|}{0.8133} & \multicolumn{1}{c|}{0.6399} & \multicolumn{1}{c|}{0.8733} & \multicolumn{1}{c|}{0.7333} & 0.8666 \\ \midrule
\multirow{2}{*}{FASHION678} & Train & \multicolumn{1}{c|}{0.9116} & \multicolumn{1}{c|}{0.3516} & \multicolumn{1}{c|}{0.8033} & \multicolumn{1}{c|}{0.7366} & \multicolumn{1}{c|}{0.8999} & \multicolumn{1}{c|}{0.8283} & 0.9283 \\ \cmidrule(l){2-9} 
 & Val & \multicolumn{1}{c|}{0.7933} & \multicolumn{1}{c|}{0.3466} & \multicolumn{1}{c|}{0.7266} & \multicolumn{1}{c|}{0.6000} & \multicolumn{1}{c|}{0.8199} & \multicolumn{1}{c|}{0.7400} & 0.8000 \\ \bottomrule
\end{tabular}
}
\vspace{-1em}
\end{table}

\section{Conclusion}
\cite{example}
We introduce knowledge distillation in QNNs using approximate synthesis to compress a pre-trained QNN for a target device or retarget a pre-trained QNN for a new device, avoiding the expensive training of new networks from scratch. Approximation error can degrade the performance of compressed or retargeted QNNs which can be largely compensated by training for a few epochs. We conduct a comprehensive numerical analysis on 7 datasets with and without noise to evaluate the efficacy of the proposed method.

\bibliographystyle{IEEEtran}
\bibliography{ref2}

\end{document}